\documentclass[preprint]{aastex61}
\usepackage{graphicx}
\usepackage{epstopdf}
\usepackage{natbib}

\submitjournal{ApJ}

\shorttitle{HSA of solar wind density fluctuations}
\shortauthors{F. Carbone et al.}

\begin{document}

\title{Arbitrary-order Hilbert spectral analysis and intermittency in solar wind density fluctuations}

\correspondingauthor{Francesco Carbone}
\email{francesco.carbone42@gmail.com, f.carbone@iia.cnr.it}

\author{Francesco Carbone}
\affil{CNR-Institute of Atmospheric Pollution Research, Division of Rende, UNICAL-Polifunzionale, 87036 Rende (CS), Italy}

\author{Luca Sorriso-Valvo}
\affiliation{CNR-Nanotec, U.O.S. di Rende, ponte P. Bucci, cubo 31C, 87036 Rende (CS), Italy}

\author{Tommaso Alberti}
\affiliation{Dipartimento di Fisica, Universit\'a della Calabria, Ponte P. Bucci, cubo 31C, 87036 Rende, Italy}

\author{Fabio Lepreti}
\affiliation{Dipartimento di Fisica, Universit\'a della Calabria, Ponte P. Bucci, cubo 31C, 87036 Rende, Italy}

\author{Christopher~H.~K. Chen}
\affiliation{School of Physics and Astronomy, Queen Mary University of London, London E1 4NS, UK}

\author{Zdenek N\v eme\v cek}
\affiliation{Faculty of Mathematics and Physics, Charles University, Prague 18000, Czech Republic}

\author{Jana \v Safr\'ankov\'a}
\affiliation{Faculty of Mathematics and Physics, Charles University, Prague 18000, Czech Republic}

\begin{abstract}

The properties of inertial and kinetic range solar wind turbulence have been investigated 
with the arbitrary-order Hilbert spectral analysis method, applied to high-resolution density measurements.
Due to the small sample size, and to the presence of strong non-stationary behavior and large-scale structures, 
the classical structure function analysis may prove to be unsuccessful in detecting the power law 
behavior in the inertial range, 
and may underestimate the scaling exponents. However, the Hilbert spectral method provides an optimal 
estimation of the scaling exponents, which have been found to be close to those for velocity fluctuations 
in fully developed hydrodynamic turbulence. 
At smaller scales, below the proton gyroscale, the system loses its intermittent multiscaling properties,
and converges to a monofractal process. The resulting scaling exponents, obtained at small scales, 
are in good agreement with those of classical fractional Brownian motion, indicating a long term memory in the process, 
and the absence of correlations around the spectral break scale.
These results provide important constraints on models of kinetic range turbulence in the solar wind.

\end{abstract}


\section{Introduction}

The solar wind is a continuous flow of plasma expanding from the solar corona 
into interplanetary space. Almost sixty years after the first spacecraft 
measurements, our knowledge of solar wind phenomena has largely advanced, 
but many aspects of the fundamental processes are still not understood. 
Among these, the properties of turbulence and its role in the non-adiabatic 
expansion of the solar wind is one of the major research goals for the 
community~\citep{Bruno2013}. 
Power-law spectra of velocity, magnetic field and density fluctuations have 
long been observed throughout the heliosphere, and represent a robust 
characteristic of solar wind turbulence~\citep{Bruno2013}. 
Unlike neutral fluid turbulence, the weakly collisional nature of solar 
wind plasma results in the presence of several scaling ranges. At scales larger 
than a few hours, the large-scale structure of the solar wind, probably of 
solar origin, generates a spectral region of energy input which can also display 
$E(\nu)\sim\nu^{-1}$ scaling~\citep{Bruno2013}, where $\nu$ is the spacecraft-frame 
measured frequency, and is characterized by mostly uncorrelated fluctuations. 
In the range between a few hours and a few seconds, the solar wind behaves as 
a magnetized flow and follows similar prescriptions to the classical Kolmogorov 
inertial range turbulence picture~\citep{K41}. Various adaptions of the 
Kolmogorov phenomenology to MHD turbulence have been 
proposed~\citep{iroshnikov64,kraichnan65,goldreich95,boldyrev06}, and the solar wind shows 
several properties that match these models, although aspects of these are still under debate.
At smaller scales, the turbulence coexists with field-particle interactions~\citep{Marsch2015}, plasma instabilities, 
and other kinetic plasma processes. In this range, a steeper power-law spectrum is generally observed, 
whose nature is still under investigation~\citep{Leamon1998,Alexandrova2008,Alexandrova2013,Chen2016,Consolini2017}.

Among the turbulence characteristics, inertial range intermittency of velocity 
and magnetic field has been deeply studied in recent years~\citep{Bruno2013}. 
Data analysis has shown that, as for neutral flows, the energy cascade is 
inhomogeneous, with the generation of localized small-scale structures which 
result in scale-dependent, non-Gaussian statistics of the field 
fluctuations~\citep{Marsch1995,Sorrisovalvo1999}. 
The appropriate estimation of the degree of intermittency is important for determining
the presence of energetic structures, such as vorticity filaments and 
current sheets, which are likely to play an important role in the dissipative 
and kinetic processes occurring in the small-scale range~\citep{Alexandrova2013}. 
For example, numerical simulations have shown that magnetic reconnection may
occur within current sheets~\citep{Servidio2012}, and plasma instabilities are also 
mostly excited in the presence of these structures~\citep{Servidio2014}.
On the other hand, the presence of intermittency in the small-scale range is still 
not fully established, although most magnetic field observations seem to indicate 
self-similar, non-intermittent scaling in this range~\citep{Kiyani2009}.
However, different techniques have given different results~\citep{Alexandrova2008}. 
This ambiguity needs to be resolved, for a better constraint on the cascade and dissipative processes. 

While most of the literature concerns the magnetic field fluctuations, some works 
have focused on the properties of density fluctuations. In particular, 
the turbulence and intermittency properties of density have been studied in the 
inertial range~\citep{Hnat2003,Telloni2014,Chen2011} and, more recently, at smaller 
scales~\citep{Chen2014,Sorrisovalvo2017}. 
The analysis of Spektr-R data have shown the presence of two power-law frequency spectra 
$E(\nu)\propto \nu^{-\beta}$, separated by a break located around the proton 
gyro-scale~\citep{Safrankova2013a,Safrankova2013b,Chen2014,Safrankova2015}. 
In particular, a scaling typical of the inertial range, characterized by a power-law 
decay of the spectral density (with a slope $\beta\simeq 5/3$) is found at large scales, 
and steeper spectra with slopes in the range $\beta\simeq 1.76$--$2.86$ exist at smaller 
scales~\citep{Sorrisovalvo2017}. 
In the inertial range, the structure functions do not show proper scaling, so that the deviation 
from $K41$~\citep{K41} was only evaluated through the standard multifractal analysis on a 
surrogate dissipation field~\citep{Sorrisovalvo2017}. 
As in the case of magnetic field, the determination of the presence of intermittency in the 
kinetic range is ambiguous, as different techniques resulted in different 
answers~\citep{Chen2014,Sorrisovalvo2017}. 
Indeed, the scaling exponents 
obtained through the structure function analysis suggested a lack of intermittency~\citep{Chen2014}, 
although also showed some variability between intervals. Similar variability was also observed in 
the multifractal spectrum~\citep{Sorrisovalvo2017}, suggesting that further analysis is required to 
fully understand the properties of this small scale dynamics.
This ambiguity motivates the use of alternative techniques, in order to understand whether or not 
nonlinear correlations are generated also in this range of scales.

The difficulty in evaluating the presence of intermittency at small scales
has several possible causes. First, the limited size of high-resolution samples
causes possible effects due to poor statistical convergence, stationarity or ergodicity. 
Second, the role of the inertial-range fluctuations may affect the statistical assessment 
of small-scale turbulence, because the presence of larger-scale strucures 
(often in the form of ramp-cliff structures as also observed in solar wind density, 
see for example the top panel of Figure~\ref{HLT}) may lead, for example, to underestimation
of the spectral index and of the structure function scaling exponents. 
Ramp-cliff structures are a common features of scalar turbulence~\citep{Shraiman2000,Warhaft2000}, 
and have been observed in a variety of turbulent shear flows in both stably and unstably 
stratified conditions~\citep{Wroblewski2007}. The typical pattern can be identified by a very 
rapid increase of the field (cliff), followed by a more gradual, or smooth decrease (ramp), 
or in reverse order~\citep{Sreenivasan1991,Celani2000,Wroblewski2007}. 
It is believed that the large scale structures may non-locally couple with the small scales through the cliff 
structure~\citep{yeung1995}. Furthermore, it has been shown that even the inertial range scaling 
may be affected by the presence of large-scale periodic forcing structures~\citep{Huang2010}.
This may have strong influence on both the small-scale and large-scale statistics~\citep{Huang2009,Huang2010}. 
Ramp-cliff structures cannot be represented by a simple monochromatic component, and
Fourier-based methods require high-order harmonic components to represent their difference. 
This lead to an asymptotic approximation process~\citep{Cohen95,Huang1998,Flandrin_book}, resulting in an artificial energy flux from large to the small scales~\citep{Huang2009}.
As a result the Fourier-based power spectrum is contaminated by this artificial energy flux, which is manifested as a shallower power spectrum~\citep{Huang2010}.
All these effects may be particularly important when the sample size is limited.

To correctly extract scaling information for solar wind proton density fluctuations, 
by minimizing the effect of the non-stationarity and the ramp-cliff structures 
embedded in the field, arbitrary-order Hilbert Spectral Analysis 
(HSA)~\citep{Huang2008,Huang2010,Carbone16} has been used in this work.
HSA formally represents an extension of classical Empirical Mode Decomposition (EMD), 
designed to characterize scale invariant properties directly in amplitude-frequency 
space~\citep{Huang2008}.
EMD was developed to process and analyze the temporal evolution of nonstationary 
data~\citep{Huang1998} and has been used in many different 
fields~\citep{Salisbury2002,Vecchio2014,Carbone_emd2016,Alberti2017}, including the analysis of 
fast, quasi-stationary solar wind high-resolution magnetic field data, 
as measured by the Cluster spacecraft~\citep{Consolini2017}. 
The main advantage of EMD is that the basis functions are derived from the signal itself. 
Since EMD analysis is adaptive (in contrast to traditional decomposition methods where the basis functions 
are fixed) and not restricted to stationary data, the data set may be analyzed 
without introducing spurious harmonics or artifacts near sharp data transitions, 
which could appear when using classical Fourier filtering or high order moments analysis.
Indeed, EMD allows local information to be extracted  through the instantaneous frequencies 
which cannot be captured by fixed-frequency methods (like Fourier or Wavelets). 
The main consequence is that the frequency is not widely spread (as for Wavelets), 
with a much better frequency definition and smaller amplitude-variation-induced frequency
modulation~\citep{Huang1998,LIU2012}.
%
%

%
%
\section{Arbitrary order Hilbert spectral analysis of solar wind proton density data}
\label{sec:spectra}
In order to perform the arbitrary-order Hilbert Analysis, 
the high-resolution solar wind proton density $n_p$ (with a sampling rate 
$\Delta t = 0.031$ sec), measured by the BMSW instrument~\citep{Safrankova2013b} on the Spektr-R 
spacecraft, have been used~\citep{Safrankova2013a,Safrankova2013b,Chen2014,Safrankova2015,Sorrisovalvo2017}.  
All of the intervals were collected during the period November 2011 to August 2012, 
and the total length of each interval is between 1 and 4 hours.  
In addition, the proton velocity $v_p$ and temperature $T_p$ were also sampled at the same  
frequency. The magnetic field $B$, not provided by Spektr-R instrumentation, 
was supplied by MFI on the Wind spacecraft, in the corresponding time interval~\citep{Chen2014}, 
and was only used for estimating the typical plasma beta $\beta_p$ of the intervals.
All of the parameters are collected in Table~\ref{table1}. More details about the data 
can be found in~\citep{Sorrisovalvo2017}. As customary, the Taylor hypothesis is used to 
shift between time and space variables via the bulk solar wind speed, which is supersonic 
and super-Alfv\'enic for all intervals. In these conditions, the time series will be used as 
an instantaneous one-dimensional cut into the turbulent flow, so that all of the arguments in terms
of space and wavevector will be given in terms of time and frequency, without loss of generality.
\begin{table*} 
	\begin{center}
		\caption{Main parameters for the eight intervals. 
			Data are from Spektr-R, except $B$, which is from the upstream Wind spacecraft. 
			The Date (dd/mm/yyyy) refers to the initial time of the measurement.
			Time is in UT, magnetic field $B$ in nT, the proton density $n_p$ in cm$^{-3}$, 
			and the proton speed $v_p$ in km s$^{-1}$.}    
		\vskip 12pt
		\begin{tabular}{ccccccc} 
			\hline
			\hline
			Interval &     Date     &        Time        & $\beta_p$ & $B$  &  $n_p$  &   $v_p$        \\ 
			\hline
			$A$      &  10/11/2011  & 15:55:40--18:46:55 &  0.78     &  4.7 &  4.6    &    370         \\
			$B$      &  01/06/2012  & 21:05:44--01:09:06 &  0.12     &  8.3 &  6.6    &    370         \\
			$C$      &  02/06/2012  & 02:34:52--03:26:43 &  0.17     &  9.1 &  7.9    &    360         \\
			$D$      &  02/06/2012  & 06:02:22--08:07:15 &  0.18     &  8.8 &  8.2    &    330         \\
			$E$      &  09/07/2012  & 08:25:56--11:09:51 &  0.06     & 12.0 &  6.0    &    400         \\
			$F$      &  09/07/2012  & 13:22:18--16:55:40 &  0.14     & 11.0 &  6.7    &    390         \\
			$G$      &  09/08/2012  & 10:48:52--15:59:13 &  0.74     &  4.7 &  4.0    &    320         \\
			$H$      &  09/08/2012  & 17:40:39--22:31:50 &  0.41     &  4.5 &  6.3    &    330         \\  
			\hline
		\end{tabular}
		\label{table1}
	\end{center}
\end{table*}

To apply HSA, the solar wind density measurements $n_p(t)$ were initially decomposed through 
classical EMD to obtain the intrinsic mode functions (IMFs), and the Hilbert transform 
was then applied to the IMFs. Within the EMD framework, the data are decomposed 
into a finite number $k$ of oscillating basis functions $\phi_j (t)$,
known as intrinsic mode functions (IMFs), characterized by an increasing time scale $\tau$,
and a residual $r_k(t)$ which describes the mean trend, if one exists, as 
\begin{equation}
n_p(t) = \sum_{j=1}^k \phi_j(t) + r_k(t) \, .
\label{emd1}
\end{equation}
The decomposition includes two stages: first, the local extrema of $n_p(t)$ are 
identified and subsequently connected through cubic spline interpolation. 
Once connected, the envelopes of local maxima and minima are obtained. 
Second, the mean $M_1 (t)$ is calculated between the two envelope functions, 
then subtracted from the original data, $h_1(t) = n_p(t) - M_1 (t)$. 
The difference $h_1 (t)$ is an IMF only if it satisfies the following criteria: 
(i) the number of local extrema and zero crossings does not differ by more than 1; 
(ii) at any point $t$, the mean value of the extrema envelopes is zero. 
When $h_1 (t)$ does not meet the above criteria, the sifting procedure is repeated 
using $h_1(t)$ as the new raw data series, and $h_{11}(t) = h_1 (t) - M_{11} (t)$ 
is generated, where $M_{11}$ (t) is the mean of the envelopes. 
The sifting procedure is repeated $m$ times until $h_{1m} (t)$ satisfies the above 
criteria. A general rule to stop the sifting is introduced by using a standard deviation 
$\sigma$, evaluated from two consecutive steps:
\begin{equation}
\sigma=\sum_{t=0}^N\frac{|h_{1(m-1)}(t)-h_{1m}(t)|^2}{h_{1(m-1)}^2(t)}\, .
\label{emd2}
\end{equation}
The iterative process stops when $\sigma$ is smaller than a threshold value 
$\sigma_\mathrm{thresh}$ \citep{Huang1998,Cummings2004}. 

\begin{figure}[h]
	\centering{\includegraphics[scale=0.47]{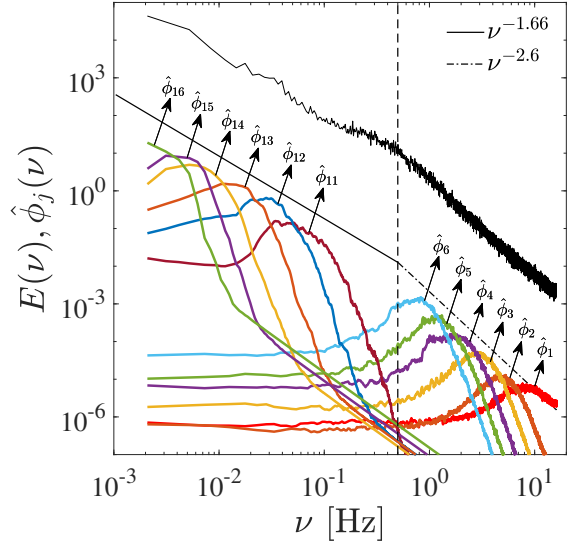}}	
	\caption{Comparison of Fourier power spectral density $E(\nu)$ (black line)
		for interval A with the Fourier power spectrum 
		of the different IMFs extracted from the EMD $\hat{\phi}(\nu)$,
		as a function of frequency $\nu$.
		The band like structure of each $\hat{\phi}(\nu)$ 
		shows the dyadic nature of the decomposition.
		The solid black line indicates $M(\nu)\sim\nu^{-1.66}$ for the inertial range,
		while the dot-dashed line indicates $M(\nu)\sim\nu^{-2.6}$ for the scales
		below the proton gyroscale.
		The vertical dashed line indicates the position of the proton gyroscale.
		The gyroscale is the frequency corresponding (via the Taylor hypothesis) to $k\rho_i=1$, $\rho_i$ being the proton gyroradius.
	}
	\label{dyadic}
\end{figure}

Since EMD acts intrinsically as a dyadic filter bank~\citep{Flandrin2004,emd_book}, 
each IMF captures a narrow spectral band in frequency 
space~\citep{Huang2008,Huang2010,Carbone16} and their superposition behaves as 
$M(\nu) \equiv \text{Max}[\phi_j(\nu)] \sim\nu^{-\alpha}$.  
In Figure~\ref{dyadic}, the results of the EMD performed on interval A
are reported. In both ranges, the behavior of $M(\nu)$ is compatible 
with the Fourier spectral indexes: $\alpha\approx 1.66$ and $\alpha\approx 2.6$, for 
the inertial range and below the proton gyroscale respectively.
%
%
%
%
%

By comparison with the Fourier spectrum, each IMF can be interpreted according
to its characteristic time scale. In particular, as visible in Figure~\ref{dyadic}, 
modes 11 to 16 capture the dynamics of the MHD inertial range, 
while modes 2 to 6 capture the small scale dynamics below the proton gyroscale. 
The intermediate range of scales, modes 7 to 10, do not show power-law scaling, 
and are representative of the dynamics across the break scale, where the Fourier spectrum 
is not described by a power-law. 
Finally, mode $\phi_1(t)$, associated to the smallest time scale, captures the 
experimental noise embedded in the datasets~\citep{Wu2004,Cummings2004}, 
setting the upper limit of the resolvable dynamics and breaking the spectral power-law decay. 
It is worth mentioning that larger scale modes with $n> 16$ also exist and are nonvanishing. 
In particular, modes 17 to 20 do not present any particular scaling, and can be 
associated with large-scale structures that could act as an energy source for the inertial range. 

Once the IMFs have been obtained, the next step of our analysis is to compute the Hilbert 
transform of each mode
\begin{equation}
\phi_j^\star = \frac{p}{\pi} \int_{-\infty}^{+\infty}\frac{\phi_j(\tau)}{t-\tau}d\tau \, ,
\label{hilbert_transf}
\end{equation}
where $p$ is the Cauchy principal value and $\phi_j(t)$ is the $j$-th IMF. 
The combination of $\phi_j(t)$ and $\phi_j^\star(t)$ defines the analytical 
signal $\mathcal{Z} = \phi_j + i\phi_j^\star = \mathcal{A}_j(t)e^{i{\theta(t)}}$, 
where $\mathcal{A}_j(t)$ is the time-dependent amplitude modulation and $\theta(t)$ 
is the phase of the mode oscillation~\citep{Cohen95}.

For each mode, the Hilbert spectrum, defined as $H(\nu,t)=\mathcal{A}^2(\nu,t)$ 
(where $\nu = d\theta/dt$ is the instantaneous frequency), provides energy 
information in the time-frequency domain. 
A marginal integration of $H(\nu,t)$ provides the Hilbert marginal spectrum 
$h(\nu)=T^{-1}\int_0^T H(\nu,t)dt$, defined as the energy density at 
frequency $\nu$~\citep{Huang1998,Huang99}. 
In addition, from the Hilbert spectrum, a joint probability density function 
$P(\nu,\mathcal{A})$ can be extracted, using the instantaneous frequency $\nu_j$ 
and the amplitude $\mathcal{A}_j$ of the $j$-th IMF. This allows the Hilbert 
marginal spectrum $h(\nu)$ to be written as
\begin{equation}
h(\nu) = \int_0^\infty P(\nu,\mathcal{A})\mathcal{A}^2 d\mathcal{A} \, ,
\label{marginal}
\end{equation} 
which corresponds to a second order statistical moment~\citep{Huang2008}.
Equation~\ref{marginal} can be generalized to the arbitrary order $q\geq 0$ by defining 
the $\nu$-dependent $q$th-order statistical moments
\begin{equation}
\mathcal{L}_q = \int_0^\infty P(\nu,\mathcal{A})\mathcal{A}^q d\mathcal{A} \, ,
\label{gen_mom}
\end{equation} 
In particular, it can be shown that $h(\nu) = \mathcal{L}_2$ represents the analogue of 
the Fourier spectral energy density~\citep{Huang2008}.
%
%
%
\begin{figure}[h]
	\centering{\includegraphics[scale=0.47]{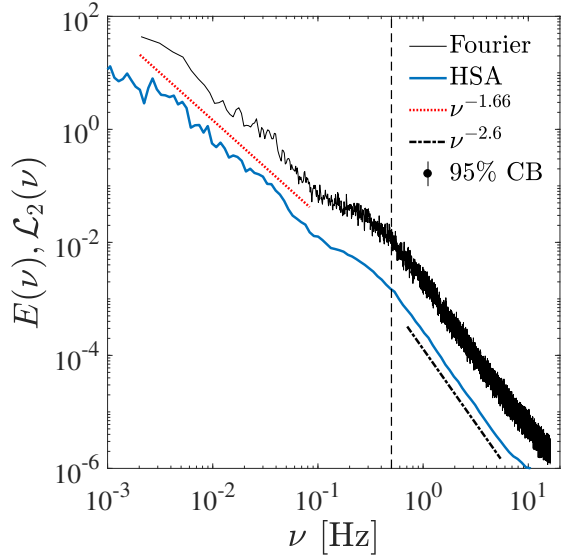}}	
	\caption{Comparison of Fourier power spectrum (thin line) with 
		$\mathcal{L}_2(\nu)$ (thick line)
		for proton density data (interval A).
		Both methods show a power-law behavior in the inertial
		range and in the small-scale range, with a slope $\beta\approx1.66$ and
		$\beta\approx2.66$ respectively. 
		The vertical dashed line represents the proton gyroscale.
		The curves have been vertically shifted for clarity.
	}
	\label{spec_comparison}
\end{figure}  

In Figure~\ref{spec_comparison}, the classical power spectral density $E(\nu)$ 
evaluated through the Fourier transform is compared with the associated $\mathcal{L}_2(\nu)$, 
obtained through the HSA. Again, the power-law behavior is present in the same two ranges 
(i.e. $\mathcal{L}_2(\nu)\sim\nu^{-\beta}$), and the slope $\beta$ is compatible with the Fourier spectrum: 
for $\mathcal{L}_2$, power-law fits give $\beta = 1.66 \pm 0.05$ for the inertial range 
and $\beta = 2.60 \pm 0.01$ for the small-scale range.
A better scaling of $\mathcal{L}_2$ can be observed at large scales, where the traditional 
Fourier spectral density shows a weak amplitude modulation, in the frequency range 
$\nu\in[10^{-2},10^{-1}]$ Hz, comparable with the typical observed size of the ramp-cliff 
structures (see Figure~\ref{HLT} top panel)~\citep{Huang2010,Carbone16}.
By analyzing the data, the ramp-cliff duration $T$ has been found in the range 
$T\in [35\pm 13, 300\pm39]$ sec, with an average duration of the order of $71.4$ sec for the ramps and $27.2$ sec for the cliffs.

Thanks to the local nature of EMD and HSA, these sources of modulation, as well as the possible
effects of ramp-cliff structures, can be constrained, isolating the properties of the cascade from the 
possible effects of the larger scale forcing and residual structures~\citep{Huang2010}.
The scaling properties of the small-scale fluctuations can thus be studied independently of the effect 
of the intermittent structures arising in the inertial range. Similarly, the inertial range scaling can be 
studied independently of the effect of of the uncorrelated large-scale fluctuations, often observed as a 
$\nu^{-1}$ spectral range~\citep{Bruno2013}. 
Due to this local nature, HSA allows a better determination of the spectral scaling exponents by mitigating 
the effects of the instrumental noise and of the larger-scale energy inhomogeneity, both in the inertial range 
and in the small-scale range. 
An exhaustive comparison of the results obtained through HSA, detrended fluctuation analysis (DFA),
structure functions (SF) and wavelet transforms (WT) has been performed in~\citep{Huang2009}.	
It was found that both the DFA and WT methods underestimate the scaling exponents, 
while the SF method may be affected by the presence of ramp-cliff structures or large scale periodic forcing~\citep{Huang2010,Carbone16}.  

%
%
%
%
\section{Intermittency Results}
\begin{figure}
	\centering{\includegraphics[scale=0.47]{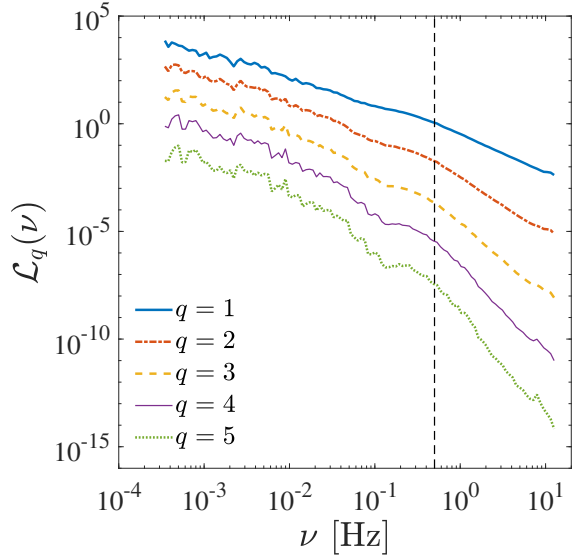}}	
	\caption{The Hilbert spectra $\mathcal{L}_q(\nu)$ for $q = 1,\dots,5$,
		obtained for interval A. The generalized spectra have been shifted for clarity. 
		The dashed line represents the proton gyroscale.}
	\label{LQ}
\end{figure}

\begin{figure*}
	\centering{\includegraphics[scale=0.6]{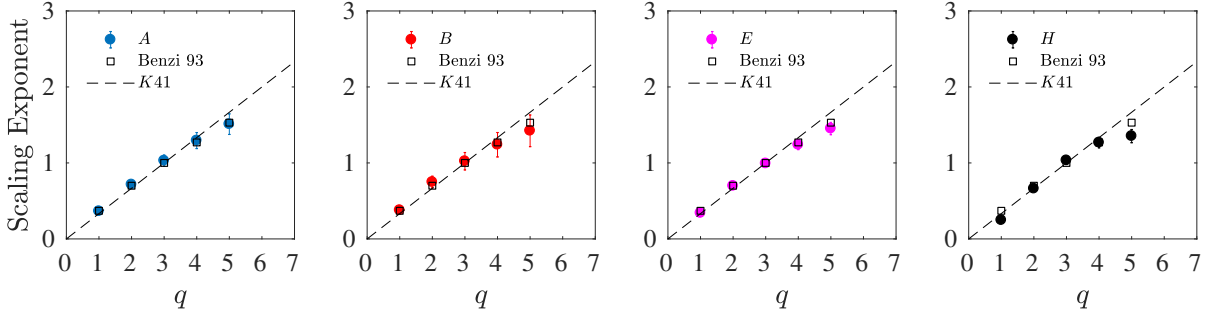}}	
	\caption{
Scaling exponents $\xi (q)$ obtained through the HSA, in the inertial range, 
for solar wind density in the intervals A, B, E, H (solid symbols); 
exponents $\zeta (q)$ obtained from velocity fluctuations measured in the 
inertial range of hydrodynamic turbulence using ESS (open symbols)~\citep{benzi93,Arneodo1996}, 
shown as reference; dashed line: theoretical expectation $q=1/3$, as estimated from dimensional 
analysis in the absence of intermittency~\citep{K41}.
}
	\label{xiq}
\end{figure*}

\begin{figure*}
	\centering{\includegraphics[scale=0.6]{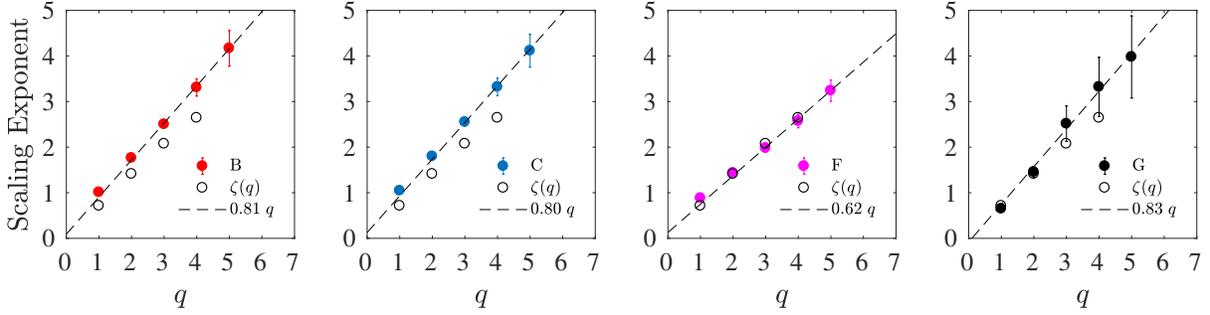}}	
	\caption{
Scaling exponents $\xi(q)$ of the solar wind density, extracted from 
intervals B, C, F, G (full symbols) at small scale. The dashed lines 
represent the fitted linear scaling $\xi(q)\simeq4/5q$ for intervals 
B, C, G, and $\xi(q)\simeq3/5q$ for interval F. The open symbols represent 
the solar wind density scaling exponents $\zeta(q)$ evaluated through the standard
Kolmogorov structure functions~\citep{Chen2014}.
	}
	\label{monofrac}
\end{figure*}

As described in Section~\ref{sec:spectra}, the generalized second-order Hilbert spectrum 
has two ranges of power-law scaling $\mathcal{L}_2\sim\nu^{-\beta}$. 
In the current Spektr-R density data $n_p(t)$, it is possible to extend the measurement of
the scaling properties of $\mathcal{L}_q(\nu)$ up to the $5$-th order. 
Figure~\ref{LQ} shows $\mathcal{L}_q(\nu)$ for orders $q = 1,\dots,5$, 
obtained from Eq.~\ref{gen_mom} using interval A. 
The resulting $\mathcal{L}_q(\nu)$ show clear scaling behavior 
$\mathcal{L}_q(\nu)\sim\nu^{-\beta_q}$ for all $q$, in the two frequency 
ranges where the spectra behave as power laws.
%
%
%
%
%
Classically, the spectral exponent $\beta$ is linked to the scaling exponent of the second order structure function 
$S(2) \equiv \langle |x(t+\tau) - x(t) |^2 \rangle \sim \tau^{\zeta(2)}$ 
(for a generic field, of component $x(t)$) 
via the relation $E(\nu)\sim \nu^{-\beta} \to \beta = 1+\zeta(2)$. 
Extending this relationship to any arbitrary order $q$, a family of generalized scaling 
exponents $\xi(q)$ can be introduced through the generalized Hilbert 
spectra~\citep{Huang2010,Carbone16} as $\xi(q)\equiv \beta_q-1$. 
The exponents $\xi(q)$ are the Hilbert analogous of the standard scaling 
exponents $\zeta(q)$ obtained through the structure functions or through the 
Extended Self-Similarity (ESS)~\citep{benzi93,Arneodo1996}. Equation~(\ref{gen_mom}) 
therefore is an alternative to the structure function scaling exponents to quantitatively 
estimate the level of intermittency in the turbulent cascade~\citep{Frish1995}, 
with the advantage of constraining the effects of noise and large-scale structure.
The scaling exponents $\xi(q)$ for the inertial range obtained from the generalized Hilbert spectra 
$\mathcal{L}_q(\nu)$ are shown in Figure~\ref{xiq}. 
The range of $\nu$ selected in order to evaluate the scaling exponent $\xi(q)$ 
lies in the closed interval $\nu \in \left[10^{-3},9\times10^{-2} \right]$ Hz. 
Intervals D and G were excluded from the analysis performed in the inertial range, as their limited size does not allow 
statistical convergence of the high order moments ($q>3$). The same figure shows a 
comparison of $\xi(q)$ with the classical exponents $\zeta(q)$ measured using Extended Self-Similarity
(ESS) for the Eulerian velocity fluctuations in fully developed hydrodynamic turbulence experiments~\citep{benzi93,Arneodo1996}. 
It is easily observed that the departure from the $K41$ scaling is captured in
the solar wind density data, and that the exponents $\xi(q)$ are similar to the standard $\zeta(q)$ 
obtained in Navier-Stokes fully developed turbulence through the ESS. 
\textbf{These are shown as reference, as neither structure functions nor the ESS analysis provided power-law
scaling for the solar wind density in this range~\citep{Chen2014}.}
%
%
%
%
%
%
%
The same analysis has been performed on the small-scale range, below the proton gyroscale,
in the range $\nu\in\left[1.5,9.8\right]$ Hz.
The scaling exponents $\xi(q)$ extracted from the generalized Hilbert spectra 
$\mathcal{L}_q(\nu)$ are shown in Figure~\ref{monofrac}. 
The results are different from the inertial range; in particular, monofractal behavior is found, 
with $\xi(q)$ presenting a linear scaling compatible with $\xi(q)\sim 4/5q$ for intervals 
A, B, C, D, G, H, and $\xi(q)\sim 3/5q$ for intervals E, F. As a comparison, in Figure~\ref{monofrac}, 
the scaling exponents $\zeta(q)$ obtained through the structure function~\citep{Chen2014} 
in a similar range of scales are reported \textbf{(the ESS analysis performed in this range gives identical 
results that are not shown for clarity)}. The difference between the two exponent sets is evident.

The weak curvature of $\zeta(q)$ could be the remnant signature of the inertial range structure, 
which acts as forcing for the dynamics in this range. The EMD-HSA analysis helps to remove these 
large-scale effects ~\citep{Huang2010}, to reveal the non-intermittent nature of the small-scale density 
fluctuations in the solar wind.
This result is in contrast with the recent multifractal analysis of the same data~\citep{Sorrisovalvo2017},
where the traditional box-counting measure applied to a surrogate dissipation field suggested a high level 
of multifractality in the small-scale range. 
However, the HSA analysis reveals the monofractal nature of the fluctuations, suggesting that the apparent 
multifractal properties may be the result of residual larger scale structure (inertial range).
A similar monofractal behavior has been found in~\citet{Consolini2017} (with different Hurst numbers $\mathcal{H}$), where the linear scaling has been obtained by analyzing the high-resolution Cluster magnetic field dataset at kinetic scales, and for each magnetic field component.

\begin{figure}[h]
	\centering{\includegraphics[scale=0.47]{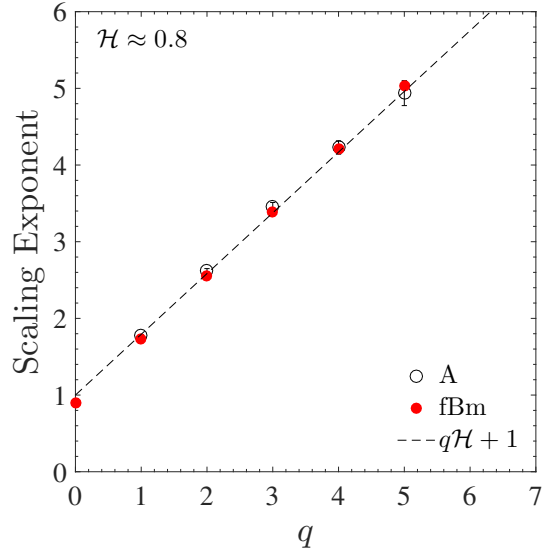}}	
	\caption{
Scaling exponents $\xi(q)+1$ extracted from interval A (open symbols), and for 
fractional Brownian motion simulation with Hurst number $\mathcal{H}=4/5$. 
The dashed line represents the theoretical scaling $q\mathcal{H}+1$.
	}
	\label{hurst}
\end{figure}

In order to further check the absence of intermittency, we compare the scaling exponents 
obtained from $n_p(t)$ with the exponents obtained from the HSA applied to fractional 
Brownian motion (fBm), with characteristic Hurst number $\mathcal{H}=4/5$ and $\mathcal{H}=3/5$,
respectively. 
The Hurst number $\mathcal{H}$ describes the long-term memory (persistence) of a process, 
or the influence ``past'' increments have on ``future'' ones.
Values in the range $\mathcal{H}\in(0.5,1]$ indicate a persistent (long-term memory, correlated) process, 
while values $\mathcal{H}\in[0,0.5)$ are associated with anti-persistent (short-term memory, anti-correlated) processes. 
$\mathcal{H} = 0.5$ indicates a completely uncorrelated process (e.g., a random walk).
In the classical Kolmogorov theory, in the absence of intermittent corrections, 
$\zeta(q)=q\mathcal{H}$. By exploiting the relation $ \xi(q) = \beta_q - 1 \propto \zeta(q)$ we expect 
$\xi(q) + 1\equiv q\mathcal{H}+1$. The comparison between $\xi(q)$ and the scaling exponents 
for the fBm ($\mathcal{H}=4/5$) is given in Figure~\ref{hurst}, which shows an excellent 
agreement supporting the absence of intermittency.
%
%
%
%
%
%
%
%

The local Hurst number has been also estimated using an alternative method. 
The evaluation of local Hurst exponent is a nontrivial issue, for which different 
approaches have been proposed in the past years. One of the most accurate, 
fast and simple methods for nonstandard, Gaussian, multi-fractional Brownian motion
is the Detrending Moving Average (DMA) technique~\citep{Alessio2002,carbone2004,Consolini2013}. 
Despite its simplicity, this method, based on the analysis of the scaling features
of the local standard deviation around a moving average, is more accurate than other methods. 
The DMA technique consists of evaluating the scaling features of the quantity:
\begin{equation}
\sigma_\mathrm{DMA}^2(n)={\frac{1}{N_\mathrm{max}-n}\sum_{j=n}^{N_\mathrm{max}}\left[f(t)-\bar{f}_n(t)\right]^2 } \, ,
\label{dma}
\end{equation}
where $\bar{f}_n(t)$ represents the average on a moving time window of length $n$, 
for different values of the time window in the interval $t_w\in[n,N_\mathrm{max}]$.
By applying this procedure, the quantity $\sigma_n(t)$ is expected to behave as 
$\sigma_n(t)\sim n^{\mathcal{H}}$.
In order to evaluate $\sigma_n(t)$ from the solar wind proton density time series, 
a moving window of approximately $N_\mathrm{max}=155$\,s has been selected. 

\begin{figure}
	\centering{\includegraphics[scale=0.33]{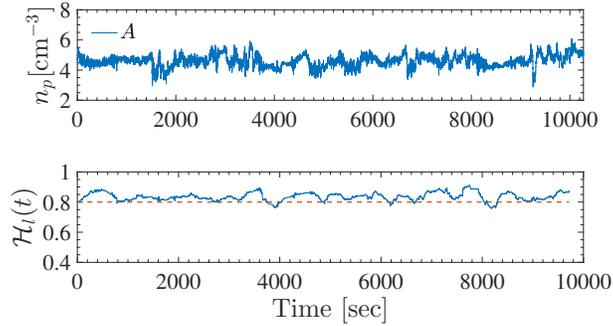}}	
	\caption{
Top panel: the solar wind proton density for interval A. The presence of 
ramp-cliff structures is visible, for example for Time between 4000 and 5000 sec. 
Bottom panel: temporal evolution of the local Hurst number $\mathcal{H}_l(t)$ 
evaluated through the DMA method (bottom panel), for interval A. 
The horizontal dashed line represents the expected value $\mathcal{H}_l = 4/5$.	
	}
	\label{HLT}
\end{figure}

The detailed temporal evolution of the small-scale $\mathcal{H}_l(t)$ is shown 
in Figure~\ref{HLT}. The top panel shows the density profile $n_p(t)$ for interval A, and the lower panel shows
the temporal evolution of the local Hurst number.
The results are in good agreement with the Hurst number extracted through the HSA, in particular
a value $\mathcal{H}\approx0.83 \pm 0.03$ has been found. The maximum percentage error with respect to the empirical
value $\mathcal{H}=0.8$ is of the order of $\Delta \mathcal{H} = 7\%$. The results relative to the other intervals are reported in Table~\ref{delta_h}

\begin{figure}[h]
	\centering{\includegraphics[scale=0.47]{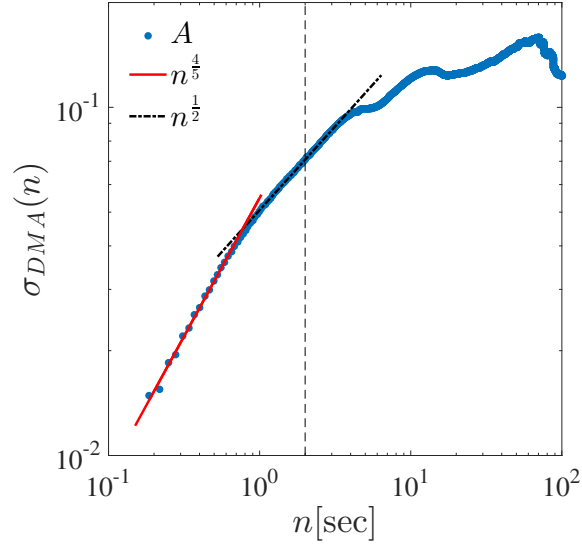}}	
	\caption{Local Hurst number evaluated through the DMA method. 
		Full symbols: $\sigma_{DMA}(n)$ as a function of the window length $n$.
		Thick red line: range of scales $n$ used to retrieve the small scale
		Hurst number $\mathcal{H}_l$. The scaling is in good agreement with
		the Hurst number evaluated through the HSA. 
		Dot-dashed line: the secondary range across the proton gyroscale.
		Vertical dashed line: the proton temporal gyroscale.}
	\label{DMA}
\end{figure}

\begin{figure}[h]
	\centering{\includegraphics[scale=0.47]{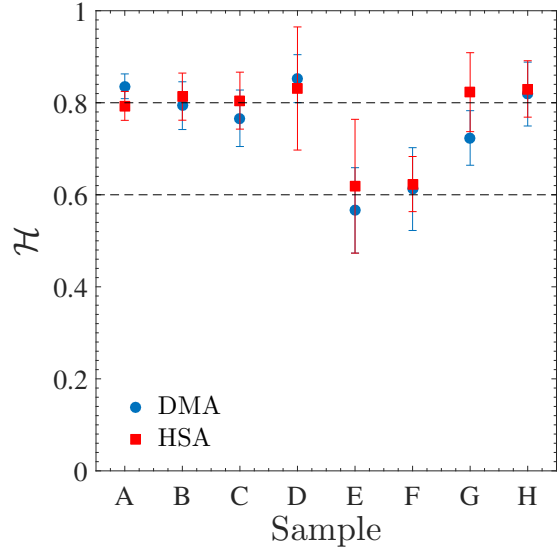}}	
	\caption{Hurst number $\mathcal{H}$, evaluated through HSA (red squares, the error bars representing the fit parameter uncertainty) 
		and through DMA (blue circles, the error being the standard deviation $\sigma_{\langle\mathcal{H}\rangle}$ from Table~\ref{delta_h}).
		The horizontal dashed lines indicate the empirical values $\mathcal{H}=4/5$ and $\mathcal{H}=3/5$.}
	\label{hsa_vs_dma}
\end{figure}

An example of $\sigma_\mathrm{DMA}(n)$, obtained from interval A at $t=6$\,s, is given 
in Figure~\ref{DMA}. At small scales $n\in[0.1,0.6]$\,s, $\sigma_\mathrm{DMA}(n)$ 
shows a good power-law scaling which provides $\mathcal{H}\approx 4/5$, in good agreement 
with HSA results. The small-scale power-law behaviour was robustly observed for all 
time windows $t_w$, in all intervals. 
A second power-law range is also always found in the intermediate range of scales 
$n\in[1,3]$\,s, around the spectral break scale, where $\mathcal{L}_2(\nu)$ 
does not show power-law scaling. 
\begin{table} 
	\begin{center}
		\caption{For all intervals (first column): the empirical estimate of the Hurst exponent $\mathcal{H}$ 
			for the scaling exponents $\xi(q)$ extracted through HSA (second column); 
			the average $\langle\mathcal{H}_l(t)\rangle$ (third column) and the standard deviation 
			$\sigma_{\langle\mathcal{H}\rangle}$ (fourth column) of the Local Hurst number evaluated 
			through the DMA method; and the maximum percentage error $\Delta \mathcal{H}\%$ with respect to 
			the empirical value $\mathcal{H}$.}   
		\vskip 12pt
		\begin{tabular}{ccccc} 
			\hline
			\hline
			Interval &  $\mathcal{H}$  & $\langle\mathcal{H}_l(t)\rangle$ & $\sigma_{\langle\mathcal{H}\rangle}$ & $\Delta \mathcal{H}\%$ \\ 
			\hline
			$A$      &  $0.8$  & $0.84$ &  $0.03$     &  $7.0$       \\
			$B$      &  $0.8$  & $0.79$ &  $0.05$     &  $5.2$          \\
			$C$      &  $0.8$  & $0.77$ &  $0.06$     &  $7.4$          \\
			$D$      &  $0.8$  & $0.85$ &  $0.05$     &  $7.4$          \\
			$E$      &  $0.6$  & $0.57$ &  $0.09$     &  $12.0$         \\
			$F$      &  $0.6$  & $0.61$ &  $0.09$     &  $12.0$          \\
			$G$      &  $0.8$  & $0.72$ &  $0.06$     &  $9.8$       \\
			$H$      &  $0.8$  & $0.82$ &  $0.07$     &  $5.4$        \\  
			\hline
		\end{tabular}
		\label{delta_h}
	\end{center}
\end{table}
It is interesting to observe that in this range the 
typical exponent for random processes $\mathcal{H}_l\approx 1/2$ is found, exposing the 
uncorrelated nature of the phenomenon during the transition between the two ranges of scales. 
For example, some mechanisms could act to decorrelate the intermittent field at the end of the 
inertial range cascade, subsequently injecting energy inhomogeneously in the small-scale range. 
The possibility of understanding the nature of the transition region dynamics using HSA analysis 
is an important issue that will be studied in depth in a dedicated work.
Finally, in the inertial range ($n>5$\,s) there is no evidence of single power-law scaling, 
in agreement with the multifractal dynamics in this range.
%
%
%
%
%
After averaging over all running windows, the mean Hurst exponents $\langle\mathcal{H}_l(t)\rangle$ 
are obtained for each interval. The results are compatible with the fit of the scaling exponents $\xi(q)$, 
as visible in Figure~\ref{hsa_vs_dma} where the two sets of values are plotted for the different intervals.
Notice that for the HSA exponents the values are consistently closer to the mean values $\mathcal{H}=4/5$
(and $\mathcal{H}=3/5$ for intervals E and F). The small discrepancy between the two techniques
could be attributed to the larger-scale structure, which introduces non-stationarity effects and artificial 
fluctuations in the scaling exponents, which may mimic multifractality~\citep{Sorrisovalvo2017}. 
Such effects are removed by HSA, so that the associated local Hurst $\mathcal{H}_l(t)$ values are less 
affected by the large-scale fluctuations.
%
%
%

\section{Conclusions}

In an attempt to describe the statistical properties of small-scale turbulence in the solar wind, 
the Empirical Mode Decomposition and the associated arbitrary-order Hilbert Spectral Analysis techniques have been 
applied for the first time to high-frequency density measurements from the Spektr-R spacecraft. 
By constructing a family of generalized Hilbert spectra $\mathcal{L}_q(\nu)$, the analogous of the 
scaling exponents of the structure functions have been evaluated from the data. 
The dyadic filter nature of EMD limits the effects of the large-scale structure~\citep{Flandrin2004,Huang2010,Carbone16}, 
allowing the identification of a scaling range corresponding to the typical inertial range of solar wind turbulence. 
Such a scaling range was not observed in this particular dataset using the traditional higher-order moments of the fluctuations.
The exponents $\xi(q)$ estimated through HSA fully capture the anomalous scaling properties related to intermittency,
exposing the multifractal nature of the inertial range turbulent cascade~\citep{Frish1995}. 
In particular, they are found to be in good agreement with the classical exponents observed in Eulerian velocity 
fluctuations in isotropic fluid turbulence~\citep{benzi93,Arneodo1996}.

The high resolution of the Spektr-R density measurements also allows the scaling properties of fluctuations 
below the proton gyroscale to be investigated, where the presence of intermittency is still debated~\citep{Alexandrova2013,Chen2014,Sorrisovalvo2017}. 
In this range, the scaling exponents obtained through HSA show a linear dependence on the order $\xi(q)\sim q\mathcal{H}$. 
This suggests that the system loses its multi-scaling properties, and converges to a non-intermittent, mono-fractal behavior. 
Two values of the Hurst number $\mathcal{H}\approx 3/5,\; 4/5$ have been found in the eight intervals under study,
indicating a persistent process (long-term memory). 
The mono-fractal nature of the small-scale fluctuations has been confirmed through a comparison with the HSA analysis 
of fractional Brownian motion with the same Hurst number. 
Furthermore, the Hurst number has also been estimated for all intervals using the Detrending Moving Average method. 
The values obtained with DMA are in good agreement with the HSA results, supporting the validity of the results.

The origin of difference between the intermittency properties of the inertial and kinetic range turbulence in the solar wind, also suggested in some previous works \citep{Kiyani2009,Chen2014}, remains an important unanswered question. The results in this paper confirm this difference, providing a more accurate measure of the scaling exponents and placing a tighter constraint on the statistical properties of the density fluctuations. Possible reasons for the difference include the increasing importance of wave-particle interactions in the kinetic range, or an inherent difference in the form of the nonlinear interactions of the cascade. The current measurements provide an important constraint on future models of kinetic turbulence.

The results presented in this paper show that the scaling properties of solar wind fluctuations need a careful analysis, 
and that the larger scale fluctuations may affect the statistical properties of the scales under study. The HSA analysis
seems to be able to reduce such effect, providing a more accurate measure of the scaling properties of the field. 

\acknowledgments{CHKC is supported by an STFC Ernest Rutherford Fellowship. JS and ZN acknowledge support of the Czech Science Foundation under Contract 16-04956S.
The authors gratefully acknowledge the anonymous
referee for his/her comments and suggestions, which contributed to the improvement of the final version of the manuscript.}

\bibliographystyle{aasjournal}

\begin{thebibliography}{}
	\expandafter\ifx\csname natexlab\endcsname\relax\def\natexlab#1{#1}\fi
	\providecommand{\url}[1]{\href{#1}{#1}}
	
	\bibitem[{Alberti {et~al.}(2017)Alberti, Consolini, Lepreti, Laurenza, Vecchio,
		\& Carbone}]{Alberti2017}
	Alberti, T., Consolini, G., Lepreti, F., {et~al.} 2017, Journal of Geophysical
	Research: Space Physics, 122, 4266
	
	\bibitem[{Alessio {et~al.}(2002)Alessio, Carbone, Castelli, \&
		Frappietro}]{Alessio2002}
	Alessio, E., Carbone, A., Castelli, G., \& Frappietro, V. 2002, The European
	Physical Journal B - Condensed Matter and Complex Systems, 27, 197
	
	\bibitem[{Alexandrova {et~al.}(2008)Alexandrova, Carbone, Veltri, \&
		Sorriso-Valvo}]{Alexandrova2008}
	Alexandrova, O., Carbone, V., Veltri, P., \& Sorriso-Valvo, L. 2008, The
	Astrophysical Journal, 674, 1153
	
	\bibitem[{{Alexandrova} {et~al.}(2013){Alexandrova}, {Chen}, {Sorriso-Valvo},
		{Horbury}, \& {Bale}}]{Alexandrova2013}
	{Alexandrova}, O., {Chen}, C.~H.~K., {Sorriso-Valvo}, L., {Horbury}, T.~S., \&
	{Bale}, S.~D. 2013, Space Sci. Rev., 178, 101
	
	\bibitem[{Arneodo {et~al.}(1996)Arneodo, Baudet, Belin, Benzi, Castaing,
		Chabaud, Chavarria, Ciliberto, Camussi, Chillà, Dubrulle, Gagne, Hebral,
		Herweijer, Marchand, Maurer, Muzy, Naert, Noullez, Peinke, Roux, Tabeling,
		Water, \& Willaime}]{Arneodo1996}
	Arneodo, A., Baudet, C., Belin, F., {et~al.} 1996, EPL (Europhysics Letters),
	34
	
	\bibitem[{Benzi {et~al.}(1993)Benzi, Ciliberto, Tripiccione, Baudet, Massaioli,
		\& Succi}]{benzi93}
	Benzi, R., Ciliberto, S., Tripiccione, R., {et~al.} 1993, Phys. Rev. E, 48, R29
	
	\bibitem[{{Boldyrev}(2006)}]{boldyrev06}
	{Boldyrev}, S. 2006, \prl, 96, 115002
	
	\bibitem[{Bruno \& Carbone(2013)}]{Bruno2013}
	Bruno, R., \& Carbone, V. 2013, Living Reviews in Solar Physics, 10, 2.
	\newblock \url{http://dx.doi.org/10.12942/lrsp-2013-2}
	
	\bibitem[{Bruno {et~al.}(2014)Bruno, Telloni, Primavera, Pietropaolo, D'Amicis,
		Sorriso-Valvo, Carbone, Malara, \& Veltri}]{Telloni2014}
	Bruno, R., Telloni, D., Primavera, L., {et~al.} 2014, The Astrophysical
	Journal, 786, 53
	
	\bibitem[{Carbone {et~al.}(2004)Carbone, Castelli, \& Stanley}]{carbone2004}
	Carbone, A., Castelli, G., \& Stanley, H. 2004, Physica A: Statistical
	Mechanics and its Applications, 344, 267
	
	\bibitem[{Carbone {et~al.}(2016{\natexlab{a}})Carbone, Gencarelli, \&
		Hedgecock}]{Carbone16}
	Carbone, F., Gencarelli, C.~N., \& Hedgecock, I.~M. 2016{\natexlab{a}}, Phys.
	Rev. E, 94, 063101
	
	\bibitem[{Carbone {et~al.}(2016{\natexlab{b}})Carbone, Landis, Gencarelli,
		Naccarato, Sprovieri, {De Simone}, Hedgecock, \& Pirrone}]{Carbone_emd2016}
	Carbone, F., Landis, M.~S., Gencarelli, C.~N., {et~al.} 2016{\natexlab{b}},
	Geophysical Research Letters, 43, 7751, 2016GL069252
	
	\bibitem[{Celani {et~al.}(2000)Celani, Lanotte, Mazzino, \&
		Vergassola}]{Celani2000}
	Celani, A., Lanotte, A., Mazzino, A., \& Vergassola, M. 2000, Phys. Rev. Lett.,
	84, 2385
	
	\bibitem[{Chen(2016)}]{Chen2016}
	Chen, C. H.~K. 2016, Journal of Plasma Physics, 82, 535820602
	
	\bibitem[{Chen {et~al.}(2011)Chen, Bale, Salem, \& Mozer}]{Chen2011}
	Chen, C. H.~K., Bale, S.~D., Salem, C., \& Mozer, F.~S. 2011, The Astrophysical
	Journal Letters, 737, L41
	
	\bibitem[{Chen {et~al.}(2014)Chen, Sorriso-Valvo, \v{S}afr\'ankov\'a, \&
		N\v{e}me\v{c}ek}]{Chen2014}
	Chen, C. H.~K., Sorriso-Valvo, L., \v{S}afr\'ankov\'a, J., \& N\v{e}me\v{c}ek,
	Z. 2014, The Astrophysical Journal Letters, 789, L8
	
	\bibitem[{Cohen(1995)}]{Cohen95}
	Cohen, L. 1995, Time-frequency analysis (Prentice Hall PTR Englewood Cliffs,
	N.J, 1995)
	
	\bibitem[{Consolini {et~al.}(2017)Consolini, Alberti, Yordanova, Marcucci, \&
		Echim}]{Consolini2017}
	Consolini, G., Alberti, T., Yordanova, E., Marcucci, M.~F., \& Echim, M. 2017,
	Journal of Physics: Conference Series, 900, 012003
	
	\bibitem[{Consolini {et~al.}(2013)Consolini, De~Marco, \&
		De~Michelis}]{Consolini2013}
	Consolini, G., De~Marco, R., \& De~Michelis, P. 2013, Nonlinear Processes in
	Geophysics, 20, 455
	
	\bibitem[{Cummings {et~al.}(2004)Cummings, Irizarry, Huang, Endy, Nisalak,
		Ungchusak, \& Burke}]{Cummings2004}
	Cummings, D.~A., Irizarry, R.~A., Huang, N.~E., {et~al.} 2004, Nature, 427, 344
	
	\bibitem[{Flandrin(1999)}]{Flandrin_book}
	Flandrin, P. 1999, Time-Frequency/Time-Scale Analysis, 1st edn., Wavelet
	Analysis and Its Applications 10 (Academic Press)
	
	\bibitem[{Flandrin {et~al.}(2004)Flandrin, Rilling, \&
		Goncalves}]{Flandrin2004}
	Flandrin, P., Rilling, G., \& Goncalves, P. 2004, IEEE Signal Processing
	Letters, 11, 112
	
	\bibitem[{Frisch(1995)}]{Frish1995}
	Frisch, U., ed. 1995, {Turbulence: the legacy of A. N. Kolmogorov} (Cambridge
	Univ. Press, Cambridge UK)
	
	\bibitem[{{Goldreich} \& {Sridhar}(1995)}]{goldreich95}
	{Goldreich}, P., \& {Sridhar}, S. 1995, \apj, 438, 763
	
	\bibitem[{He {et~al.}(2015)He, Wang, Tu, Marsch, \& Zong}]{Marsch2015}
	He, J., Wang, L., Tu, C., Marsch, E., \& Zong, Q. 2015, The Astrophysical
	Journal Letters, 800, L31
	
	\bibitem[{Hnat {et~al.}(2003)Hnat, Chapman, \& Rowlands}]{Hnat2003}
	Hnat, B., Chapman, S.~C., \& Rowlands, G. 2003, Phys. Rev. E, 67, 056404
	
	\bibitem[{Huang \& Shen(2005)}]{emd_book}
	Huang, N.~E., \& Shen, S. S.~P., eds. 2005, {The Hilbert-Huang Transform and
		Its Applications} (World Scientific, Singapore)
	
	\bibitem[{Huang {et~al.}(1999)Huang, Shen, \& Long}]{Huang99}
	Huang, N.~E., Shen, Z., \& Long, S.~R. 1999, Annual Review of Fluid Mechanics,
	31
	
	\bibitem[{Huang {et~al.}(1998)Huang, Shen, Long, Wu, Shih, Zheng, Yen, Tung, \&
		Liu}]{Huang1998}
	Huang, N.~E., Shen, Z., Long, S.~R., {et~al.} 1998, Proceedings of the Royal
	Society of London A: Mathematical, Physical and Engineering Sciences, 454,
	903
	
	\bibitem[{Huang {et~al.}(2011)Huang, Schmitt, Hermand, Gagne, Lu, \&
		Liu}]{Huang2009}
	Huang, Y.~X., Schmitt, F.~G., Hermand, J.-P., {et~al.} 2011, Phys. Rev. E, 84,
	016208
	
	\bibitem[{Huang {et~al.}(2010)Huang, Schmitt, Lu, Fougairolles, Gagne, \&
		Liu}]{Huang2010}
	Huang, Y.~X., Schmitt, F.~G., Lu, Z.~M., {et~al.} 2010, Phys. Rev. E, 82,
	026319
	
	\bibitem[{Huang {et~al.}(2008)Huang, Schmitt, Lu, \& Liu}]{Huang2008}
	Huang, Y.~X., Schmitt, F.~G., Lu, Z.~M., \& Liu, Y.~L. 2008, EPL (Europhysics
	Letters), 84, 40010
	
	\bibitem[{{Iroshnikov}(1964)}]{iroshnikov64}
	{Iroshnikov}, P.~S. 1964, Soviet Astron., 7, 566
	
	\bibitem[{Kiyani {et~al.}(2009)Kiyani, Chapman, Khotyaintsev, Dunlop, \&
		Sahraoui}]{Kiyani2009}
	Kiyani, K.~H., Chapman, S.~C., Khotyaintsev, Y.~V., Dunlop, M.~W., \& Sahraoui,
	F. 2009, Phys. Rev. Lett., 103, 075006
	
	\bibitem[{Kolmogorov(1941)}]{K41}
	Kolmogorov, A.~N. 1941, C. R. Acad. Sci. U.R.S.S, 36, 301
	
	\bibitem[{{Kraichnan}(1965)}]{kraichnan65}
	{Kraichnan}, R.~H. 1965, Phys. Fluids, 8, 1385
	
	\bibitem[{Leamon {et~al.}(1998)Leamon, Smith, Ness, Matthaeus, \&
		Wong}]{Leamon1998}
	Leamon, R.~J., Smith, C.~W., Ness, N.~F., Matthaeus, W.~H., \& Wong, H.~K.
	1998, Journal of Geophysical Research: Space Physics, 103, 4775
	
	\bibitem[{Liu {et~al.}(2012)Liu, Fujita, Watanabe, \& Mitani}]{LIU2012}
	Liu, Q., Fujita, T., Watanabe, M., \& Mitani, Y. 2012, IFAC Proceedings
	Volumes, 45, 144 , 8th Power Plant and Power System Control Symposium
	
	\bibitem[{Marsch \& Tu(1997)}]{Marsch1995}
	Marsch, E., \& Tu, C.-Y. 1997, Nonlinear Processes in Geophysics, 4, 101
	
	\bibitem[{{\v{S}}afr{\'a}nkov{\'a} {et~al.}(2013){\v{S}}afr{\'a}nkov{\'a},
		N{\v{e}}me{\v{c}}ek, P{\v{r}}ech, Zastenker, {\v{C}}erm{\'a}k, Chesalin,
		Kom{\'a}rek, Vaverka, Ber{\'a}nek, Pavl{\r{u}}, Gavrilova, Karimov, \&
		Leibov}]{Safrankova2013b}
	{\v{S}}afr{\'a}nkov{\'a}, J., N{\v{e}}me{\v{c}}ek, Z., P{\v{r}}ech, L.,
	{et~al.} 2013, Space Science Reviews, 175, 165
	
	\bibitem[{Salisbury \& Wimbush(2002)}]{Salisbury2002}
	Salisbury, J.~I., \& Wimbush, M. 2002, Nonlinear Processes in Geophysics, 9,
	341
	
	\bibitem[{Servidio {et~al.}(2014)Servidio, Osman, Valentini, Perrone, Califano,
		Chapman, Matthaeus, \& Veltri}]{Servidio2014}
	Servidio, S., Osman, K.~T., Valentini, F., {et~al.} 2014, The Astrophysical
	Journal Letters, 781, L27
	
	\bibitem[{Servidio {et~al.}(2012)Servidio, Valentini, Califano, \&
		Veltri}]{Servidio2012}
	Servidio, S., Valentini, F., Califano, F., \& Veltri, P. 2012, Phys. Rev.
	Lett., 108, 045001
	
	\bibitem[{Shraiman(2000)}]{Shraiman2000}
	Shraiman, Boris I.;~Siggia, E.~D. 2000, Nature, 405, doi:10.1038/35015000
	
	\bibitem[{Sorriso-Valvo {et~al.}(2017)Sorriso-Valvo, Carbone, Leonardis, Chen,
		\v{S}afr\'ankov\'a, \& N\v{e}me\v{c}ek}]{Sorrisovalvo2017}
	Sorriso-Valvo, L., Carbone, F., Leonardis, E., {et~al.} 2017, Advances in Space
	Research, 59, 1642
	
	\bibitem[{Sorriso-Valvo {et~al.}(1999)Sorriso-Valvo, Carbone, Veltri,
		Consolini, \& Bruno}]{Sorrisovalvo1999}
	Sorriso-Valvo, L., Carbone, V., Veltri, P., Consolini, G., \& Bruno, R. 1999,
	Geophysical Research Letters, 26, 1801
	
	\bibitem[{Sreenivasan(1991)}]{Sreenivasan1991}
	Sreenivasan, K.~R. 1991, Proceedings of the Royal Society of London A:
	Mathematical, Physical and Engineering Sciences, 434, 165
	
	\bibitem[{Vecchio {et~al.}(2014)Vecchio, Anzidei, \& Carbone}]{Vecchio2014}
	Vecchio, A., Anzidei, M., \& Carbone, V. 2014, Journal of Geodynamics, 79, 39
	
	\bibitem[{\v{S}afr\'ankov\'a {et~al.}(2015)\v{S}afr\'ankov\'a, N\v{e}me\v{c}ek,
		N\v{e}mec, P\v{r}ech, Pit\v{n}a, Chen, \& Zastenker}]{Safrankova2015}
	\v{S}afr\'ankov\'a, J., N\v{e}me\v{c}ek, Z., N\v{e}mec, F., {et~al.} 2015, The
	Astrophysical Journal, 803, 107
	
	\bibitem[{\v{S}afr\'ankov\'a {et~al.}(2013)\v{S}afr\'ankov\'a, N\v{e}me\v{c}ek,
		P~\v{r}ech, \& Zastenker}]{Safrankova2013a}
	\v{S}afr\'ankov\'a, J., N\v{e}me\v{c}ek, Z.~v., P~\v{r}ech, L., \& Zastenker,
	G.~N. 2013, Phys. Rev. Lett., 110, 025004
	
	\bibitem[{Warhaft(2000)}]{Warhaft2000}
	Warhaft, Z. 2000, Annual Review of Fluid Mechanics, 32, 203
	
	\bibitem[{Wroblewski {et~al.}(2007)Wroblewski, Coté, Hacker, \&
		Dobosy}]{Wroblewski2007}
	Wroblewski, D.~E., Coté, O.~R., Hacker, J.~M., \& Dobosy, R.~J. 2007, Journal
	of the Atmospheric Sciences, 64, 2521
	
	\bibitem[{Wu \& Huang(2004)}]{Wu2004}
	Wu, Z., \& Huang, N.~E. 2004, Proceedings of the Royal Society of London A:
	Mathematical, Physical and Engineering Sciences, 460, 1597
	
	\bibitem[{Yeung {et~al.}(1995)Yeung, Brasseur, \& Wang}]{yeung1995}
	Yeung, P.~K., Brasseur, J.~G., \& Wang, Q. 1995, Journal of Fluid Mechanics,
	283, 43–95
	
\end{thebibliography}

\end{document}